\documentclass[a4paper,12pt]{article}

\usepackage{amssymb, amsmath}
\usepackage[pdftex]{graphicx}
\graphicspath{{Convert/}}
\usepackage{cite}
\def\be{\begin{equation}}
\def\ee{\end{equation}}
\def\bea{\begin{eqnarray}}
\def\eea{\end{eqnarray}}

\renewcommand{\baselinestretch}{1.30}
\title{ {\bf Probabilistic and Geometric Languages \\ in the Context of \\ the Principle of Least Action}}

\author{Vladislav E. Terekhovich \\  \small \sl  Department of Philosophy of Science and Techniques, \\ \small \sl  Faculty of Philosophy, St. Petersburg State University, St. Petersburg, Russia \\ \small \sl 
E-mail:  v.terekhovich@gmail.com}

\date{}

\begin{document}

\maketitle

{
\renewcommand{\baselinestretch}{1.0}
\selectfont
\begin{abstract}
\tolerance = 400
This paper explores the issue of the unification of the three languages of physics, the geometric language of forces, geometric language of fields or 4-dimensional space-time, and probabilistic language of quantum mechanics. On the one hand, equations in each language may be derived from the Principle of Least Action (PLA). On the other hand, Feynman's path integral method could explain the physical meaning of PLA. The axioms of\,classical and relativistic mechanics can be considered as consequences of Feynman's formulation of quantum mechanics. 

\end{abstract}
}

\section {Introduction}

We are used to describing each area of nature within the framework of a specific branches of science, which uses special methods or languages. Under the notion language of the branch of science I mean not only a formal set of definitions, axioms, logical rules, and mathematical tools. The language also includes a view of causality and reality\footnote{V. Heisenberg wrote about the language of science as a set of concepts, logic, and ontology axioms \cite{Heisenberg2004}. T. Kuhn said about the languages used in science and included many assumptions about nature \cite{Kuhn2003}}. 

It is well accepted that the different languages coexist in physics. They are based on the concepts of forces, fields, streams, stability, space-time geometry, statistics, probability, and others. For instance, a motion of the macroscopic objects is described in terms of the forces or fields. The language of curved space-time is used for the description of the cosmological objects. The geometric representation of the objects and deterministic causality unite both these languages. However, philosophical foundations of the languages are different. Until the present, many scientists have believed that the probabilistic language of thermodynamics, especially of non-equilibrium, is a statistical approximation of classical mechanics. After the Copenhagen formulation of quantum mechanics (QM), some physicists accepted that geometrical and deterministic causality are not applicable to the micro objects. Some of them, like Max Born, are sure that the wave theory must dispose of the means of translating its results into the language of the ordinary objects of mechanics \cite{Born1963a}. All attempts to reduce quantum probability to statistics or to consider the probabilistic description as incomplete have failed. Many interpretations of the physical meaning of QM have appeared \cite{Jammer1974, Bohm1980, Zurek2003, Sevalnikov2003, GreensteinZajonc2006, Penrose2007}. However, it remains unclear how to reconcile the classical laws of nature with impossibility of the deterministic definition of the quantum events in time and space. This paper attempts to give a partial answer to this question.

Richard Feynman believed that every decent physicist-theorist knows six or seven theories that describe of the same physical facts \cite{Feynman1987b}. It is known that the philosophical foundations of such theories often contradict each other. Scientists do not like this fact, at least for aesthetic reasons. They understand that scientific knowledge equals an awareness of connections \cite{Heisenberg2004a}. The connections are realized within the language of the scientific community, which is, in turn, connected with the dominant paradigm \cite{Kuhn2003}. Thus, mutual understanding is defined by the common language.

In this paper, I describe how to remove some contradictions between the geometrical descriptions in terms of forces, fields, and 4-dimensional space-time and the probabilistic laws of QM.  My proposal is based on the variational principle---the Principle of Least Action (PLA). The physical and philosophical meaning of it is disclosed by means of Feynman's formulation of QM using the path integral or many paths method. 

It is considered that the axioms of classical mechanics, classical field theory, and general relativity are based on the happy guesses of their creators (that is not the entire truth). I consider these theories as the necessary consequences of QM. The equations of the main fields of physics can be represented as the limit of quantum equations. This is because the geometric description of motion in\,n-dimensional space can be represented as a convenient mathematical approximation of more fundamental probabilistic descriptions.

PLA was formulated by Maupertuis in 1744. Euler gave it a mathematical form, Lagrange, D'Alamber, Hamilton, Gauss, Helmholtz, and others took part in its improvement. Einstein considered that whole general relativity could be derived from this single variational principle \cite{Einstein1965}. Planck named it as a more universal law of nature than the law of conservation of energy and momentum, so PLA "dominates above all reversible phenomena of physics" \cite{Planck1925}. Eddington wrote about two great generalizations of science: PLA and the second law of thermodynamics \cite{Eddington2003}. Moore states that "this principle lies at the core of much of contemporary theoretical physics" \cite{Moore1996}. Attention to PLA has not weakened, especially in connection with quantum physics \cite{HancTulejaHancova2003, Taylor2003, GrayKarlNov2004, OgbornTaylor2005, GrayTaylor2007, Dyson2007, Sbitnev2008}, and cosmology \cite{TaylorWheeler1992, TaylorWheeler2000, NusserBranchini2000, Marchal2002, GrayPoisson2010}.

To prove my approach, I show that the basic equations of some physical theories are equivalent to one of PLA's forms, which are equivalent to each other. I show that each  form of PLA could be represented as the limit of Feynman's path integral method based on the notion of quantum probability amplitudes. 

\noindent 

\section{Four methods to describe motion of body and their philosophical foundations}

There are four methods to predict the flight path of a body thrown angularly to the horizon (see \cite{Feynman2004a, Feynman1987}).

\textit{The first method.} Newtonian theory says that the body has inertia and is attracted by the Earth with a certain force. The forces of inertia and gravity depend on the body's weight. The actual movement at each moment is a sum of movements caused by both forces. According to Newton's idea, the body has a mysterious internal "tendency" of moving straight with constant speed. If the body "feels" the effect of external forces, it is accelerated. It is assumed that the force effect is felt at a distance (not locally) and depend on the body's height from the center of the Earth. If body's initial position and the vector of speed are known, we can write down an equation to calculate all points of its trajectory. As a result, the actual trajectory is defined as the sum of two virtual trajectories---the horizontal and vertical.

\textit{The second method.} If we do not like the "mystical" effect at a distance, we can describe the same body's flight in terms of field theory. The field is a collection of numbers at each point of space. These numbers, called "potentials", vary from one point of space to another. If we put the body at any point of space, we find the force acting on the body in the direction, in which the potential decreases most rapidly. In other words, this force is proportional to the speed of the potential decrease, or the vector of the force is an antigradient of potential energy. The actual body's trajectory is determined by the force at each point in space.

It seems that the body "probes" space along all virtual trajectories around itself and rushes along single trajectory where the potential of the gravitational field is minimum. The faster the potential decreases, the faster the body rushes (ones usually say that the force acting on the body is greater). The field formulation allows us to predict the body's flight, if we know what is happening in the present moment at each point around it. The clause "at the present moment" is important, because virtual "probes" of space do not take any real time. Unfortunately, without metaphors we cannot explain how the body "learns" the value of the potential at the neighbouring points.

\textit{The third method.} Another method of predicting the body's flight is very different from the first or the second ones, especially in a philosophical sense. It is not necessary to know what is happening at the close moment in time or at the neighbouring points of space. We only need to know the body's initial and final positions in space and time. PLA states that the actual body's trajectory from one point to another in the same time is the one from all possible ones, for which a functional called "action" is minimum or stationary. Hamilton's form of PLA\footnote{This principle arose from the optical-mechanical analogy with Fermat's principle, by which the light moves along the path that takes less time. Schr\"odinger in his Nobel speech showed that only in terms of the wave method of observation do Hamilton's and Fermat's principles open their true value \cite{Schroedinger1934}.} says that along the actual body's trajectory the difference between its average kinetic and potential energies reaches a minimum in comparison with all possible trajectories. The differential equations of the body's motion in the gravitational field (Euler-Lagrange equations) could be derived from PLA \cite{Landau2004}. Each virtual trajectory of the body corresponds to a certain amount of the action, but only that trajectory is actual, for which the action is minimum. Only this trajectory is observed as real and exactly coincides with the results of two previous methods. Now we do not need to think about any forces. We also do not need for a fictitious inertial force, because in the absence of the potential field, the body's trajectory with the least action is the straight line with constant speed\footnote{PLA has an advantage over the principle of conservation of energy and variational principles of mechanics (D'Alamber's, virtual displacements, Gauss', and others), because in one equation, PLA gives the relation between the values of space, time, and potentials \cite{Planck1911}.}.

\textit{The fourth method.} In general relativity theory, there are no attractive forces and potential fields of gravity. Instead, overall geometric space-time is curved under the influence of Earth's mass. The body moves inertially along a world line (called a geodesic) in space-time between initial and final events. The form of the geodesic is calculated by the equation for 4-dimensional space-time. For Earth's conditions the result of calculations coincides with the results of the previous three methods, and the form of the geodesic is accurately described by PLA of classical mechanics \cite{Lanczos1965a}. For a free body, the actual world line between two events is the one, from all possible world lines, for which a value of the body's proper time is maximum or stationary. This line is the geodesic. This principle is called the Principle of Maximum Proper Time\footnote{Using the Principle of Maximal Aging, we can study stars and black holes without the tensors and field equations of general relativity (see \cite{TaylorWheeler1992, TaylorWheeler2000}).}. For weak gravitational fields and low speeds, it is reduced to PLA in Hamilton's form (see the third method).\textbf{}

\noindent 

\section{Feynman's formulation of quantum mechanics}

\noindent 
\looseness=-1
In 1942, Richard Feynman \cite{Feynman2005} used the ideas of Huygens and Fresnel, which had formerly inspired Schr\"odinger to his wave equation, and proposed a new formulation of QM. He replaced a classical concept of a particle's motion along a single and unique path by a representation of the motion along an infinite set of conceivable paths, and mathematically described it by a functional integral. He assumed that the particle moves simultaneously along all possible paths, each of these is associated with a quantum amplitude of probability. The quantum amplitudes of all paths are extinguished at the final point, so that the maximum probability corresponds to the actual path, for which the variation of some functional is zero. Feynman called this functional "the action" by analogy with classical mechanics and connected it with the quantum phase of waves of probability \cite{Feynman1968}. Every possible path of the particle possesses the phase, and the amplitudes near the actual path are nearly in the same phase. Thus, they reinforce each other and generate significant effects, observable as "real". Other paths exist too (they are called virtual or imaginable), but they are not observed or, more precisely, their probability to be observed are very small. It could be called as "probabilistic~existence". The probability of observing is given by the square of a modulus of the amplitude (wave function). This formulation of QM is mathematically equivalent to Heisenberg's matrix method and Schr\"odinger's wave equation \cite{OgbornHancTaylor2006}.

\noindent 

\section{Classical body and quantum mechanics}

\noindent 

Consider how QM relates to the flight of the classical body. The classical laws are deterministic, they accurately predict the body's behaviour, and, it seems, they are not connected with the probability of the micro objects. Nevertheless, Feynman concluded that QM is more primary than classical mechanics and general relativity, as far as the fundamental laws of physics can be expressed in the form of PLA \cite{Feynman2004}. Even the relationship between symmetry and the laws of conservation, as articulated in Noether's theorem is based on PLA, which follows from the laws of quantum mechanics \cite{Feynman1987a}.\textbf{}

According to Feynman, a classical body, as well as a photon or an electron moves simultaneously along all possible paths or world lines between initial and final events. As the phase of quantum amplitude is very high, a set of world lines that makes a significant contribution to the probability of the body's detection, reduces to a narrow bundle. In the limit it contracts to the single world line predicted by the Hamilton's classical form of PLA \cite{Taylor2003}. It is like the third method in Section 2. 

What\,Newtonian physics treats as cause and effect (the force producing acceleration), the quantum "many paths" view treats as a balance of changes in phase produced by changes in kinetic and potential energy \cite{OgbornTaylor2005}. So classical mechanics and field theory become short-wave approximations of QM, and the action is given the meaning of the phase of quantum amplitude. It is no longer necessary to use the concept of the forces acting on the body. It is enough that the body simultaneously "passes" along all possible paths from one point to another and "selects" the path, for which the action is minimum \cite{Feynman2004b}. Perhaps, the term "select" is superfluous in this case, because the classical trajectory is not selected by the body, but by the rule of addition of the quantum phases.

\begin{figure}[!ht]
\centering
\includegraphics[scale=0.52]{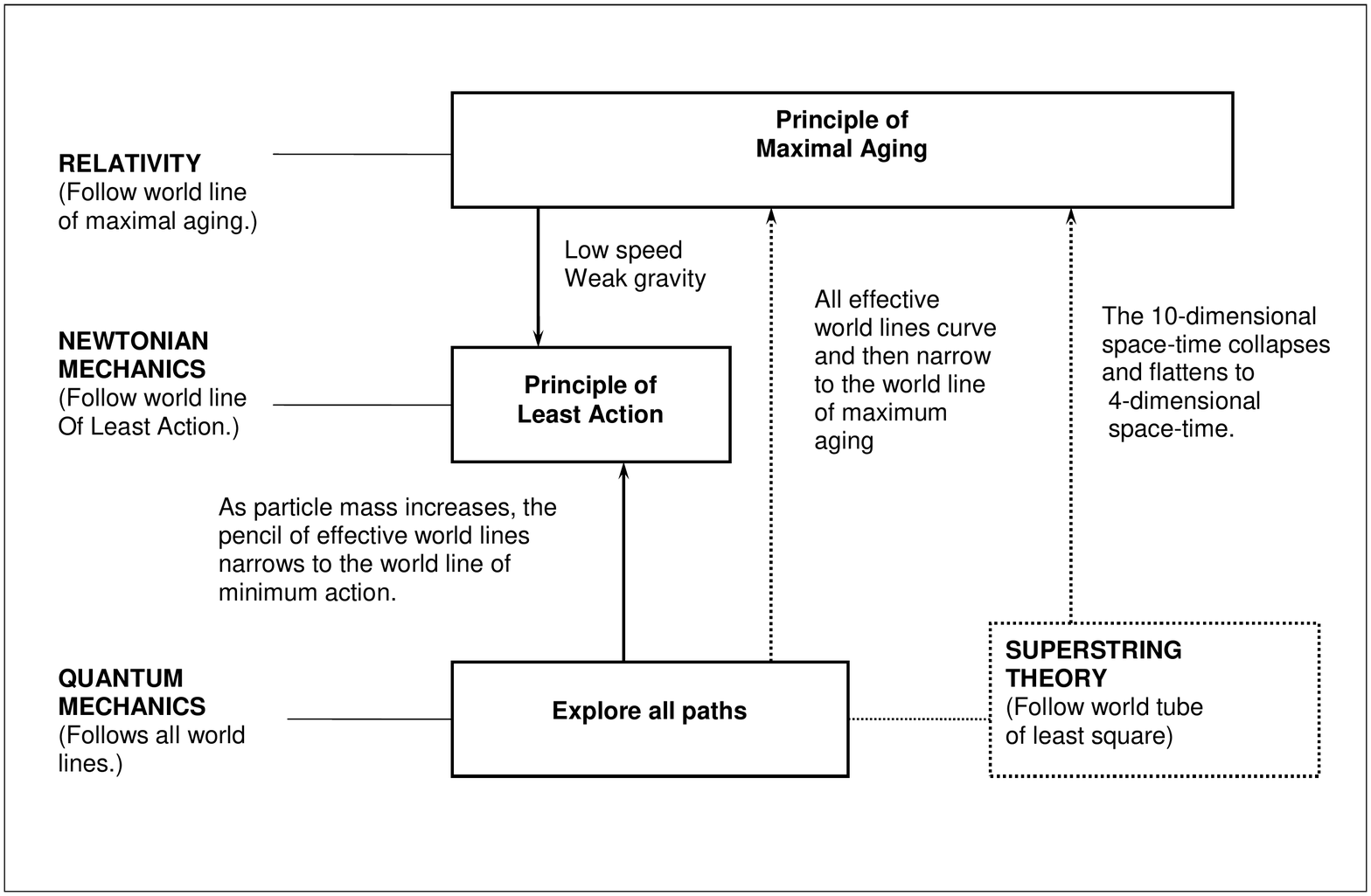}
\caption{\footnotesize Story line showing the principle of least action sandwiched between relativity and quantum mechanics \cite{Taylor2003}. In addition relativity becomes the limiting case and approximation of quantum mechanics.}
\label{pict:1}
\end{figure}

\looseness=-1
According to Taylor's figurative expression \cite{Taylor2003}, a stone moving with nonrelativistic speed in the region of a small space-time curvature obeys nature's command: \textit{Follow the path of least action}! The stone moving with any possible speed in curved space-time obeys nature's command: \textit{Follow the path of maximum aging (or maximum proper time)}! The electron obeys nature's command: \textit{Explore all paths}! \cite{Taylor2003}. Taylor proposes a scheme where PLA, on the one hand, is a limiting case of the Principle of Maximal Aging, on the other hand, a limiting case of Feynman's principle "Explore all paths". In other words, Newtonian mechanics becomes a limiting case and approximation of general relativity and QM at the same time. I suggest ex\-te\-nd\-ing Taylor's scheme using his metaphors. My additions are indicated by dashed lines (Fig.~\ref{pict:1}).

Firstly, the Principle of Maximum Aging could be considered as the limiting case of Feynman's principle "Explore all paths" for strong gravity. Under the influence of massive objects the pencil of the effective world lines of the particle curves. The phases of all amplitudes added together, so possible world lines reduce to one world line of maximum aging. Secondly, we could apply nature's command, "Explore all paths", for the space-time of any dimensions and curvature. The Principle of Maximum Aging is applied only for the smooth\,4-dimensional space-time. The use of the formalism of QM for such space-time creates an infinite and even negative probability inevitably. One of the mathematical solutions of this problem is offered by a theory of superstrings, which requires extra dimensions \cite{Greene2005}. According to this theory, in each point of 4-dimensional space-time, there are six or more extra collapsed dimensions. If the superstring theory is true (there is no evidence of this), we can assume that the Principle of Maximum Aging is also an approximation of the path integral method. When the scale increases, the n-dimensional space-time collapses and flattens into \,4-dimensional one\footnote{It is assumed that the string moves in space along the world sheet or world tube. To calculate the trajectory of its movement, we should minimize the analog of path's length--area of tube \cite{Gross2006}.}. All possible paths of micro objects are stable only in\,4-dimensional space-time, therefore, possible paths in 10-dimensional space-time reduce to the possible paths in 4-dimensional space-time due to interference. So as well as classical mechanics, general relativity also becomes the limiting case and approximation of QM.

\noindent 

\section{Discussion}

\noindent 

We have attempted to combine the languages of physics by means of PLA, in spite of the philosophical status of PLA is unclear. One of the reasons is a phantom of final cause. The explanation of PLA by the simplicity and perfection of nature in a teleological sence does not coordinate with any scientific paradigms. Gradually, PLA has turned into a pure heuristic rule. The opponents of philosophical interpretations of PLA were D'Alamber, Lagrange, Jacobi, Einstein, Prigogine, and others. Mach found that the variational principles of mechanics are no more than other mathematical formulations of Newtonian laws and that they do not contain anything new. However, he added that modern mathematics did not provide any other method to formulate a covariant and at the same time a compatible system of field equations \cite{Lanczos1965}.

Born wrote that Einstein's law of gravity, which includes Newtonian laws as the limiting case, could also be derived from PLA. Following Mach, Born emphasized that extreme descriptions talk not about properties of nature but about our aspiration for economy of thinking \cite{Born1963}. According to other opinions, PLA does not have only methodological meaning, but expresses the unity and interconnection of symmetry, the conservation principles, and causality \cite{Razumovsky1975}. PLA does not summarize only the physical causality, but also regularity, necessity, probability, and connection of states \cite{Asseev1977}.  The law of conservation of energy, as well as other laws of conservation can be derived from the action and variational principles \cite{Goldstein2002, Brizard2008}. It is believed that QM in Feynman's form appears as the generalization of classical mechanics \cite{Myakishev1973}, and the application of path integral provides a clear and elegant language, which describes the transition from classical to quantum physics \cite{Ramon1984}.

Following Euler, Lagrange, and Hamilton, the creators of QM borrowed an optical-mechanical analogy from geometrical optics (Fermat's and Huygens's principles) and mechanics (Hamilton's principle) \cite{Schroedinger1959}. Hilbert and Einstein used the same analogy when they wrote their equations of general relativity \cite{Einstein1959}. Perhaps, methodological convenience is not the sole reason for this analogy. I think that the use of the same analogy in the different languages points at their common essence. 

General relativity is able to unite Newtonian mechanics and the field theory of motions in 4-dimensional space-time with any curvature. However, this language is not applied to Planck's scale. Only one language successfully works at three levels, it is the language of PLA. At the level of curved 4-dimensional space-time, Einstein's equations are equivalent to the Principle of Maximum Aging for free particles and PLA for gravitational fields. At the classical level, Newtonian and field equations are equivalent to PLA of free bodies or fields of different types \cite{Landau2003}. At the quantum level, the field's equations are equivalent to the path integral method. The last one, in its turn, explains why PLA works at all levels. So just the method of Feynman could answer why quantum, classical, and relativistic objects obey the same principles. 

Physics is an amazing science. The same observed result could be obtained within the framework of four languages with close mathematical precision. Each language is based on different logical and philosophical grounds. Which one is correct? I think, this is not the right question, and every language is correct in its own field of nature. We should formulate another goal---to find a method of combining the classical and relativistic geometric languages with the probabilistic language of QM. I assume that it could be the new language based on PLA and Feynman's formulation of QM. It has some advantages.

\begin{enumerate}
\item The basic physical theories can be represented as approximations and limiting cases of this language. We do not need the concept of "force", replacing it with changing phases of quantum amplitudes.
\item This language accounts for the transition from probabilistic to deterministic causality. It is enough to connect the minimal principles with the concept of probability.
\item Results predicted by this language correspond to observations of micro, macro and mega objects, for any speeds and dimensions of space.
\item This language is based on the simple set of notions; it has simple and universal mathematical tools---the calculus of variations.
\end{enumerate}

Of course, there are some difficulties with this language. The path integral method has some problems in the quantum field theory. It is unexplained why PLA in each field of physics has very different forms. We do not understand the similarities between all forms of the action. In classical mechanics, the action is the difference between average kinetic and average potential energy. In general relativity, the action is the proper time. In QM, the action is the probability amplitude. There are other questions as well. Why is the action always extremal? Why is any form of the action invariant concerning transformations of space-time? How is the action connected with energy, space, and time?\textbf{}

However, the main problem of new language is philosophical. Our common sense protests against the proposed explanation of the essence of phenomena. If PLA is not the convenient method, and path integral is not only the useful metaphor, as most physicists belive, then how is it possible that everyday objects locate simultaneously at different points of space-time? The pfysicists say that it happens virtually, but they do not explain what it means. Does it happen, really or not? The most radical idea of the language based on PLA and the path integral method is that any classical objects "explore" all possible paths as well as the quantum particles. Due to interference of their possible paths, classical objects are found in the state or on the path corresponding to the minimum action. What does the word "explore" really mean, when applied to inanimate matter? Feynman did not point at any philosophical sense of his method, considering it only as a convenient formalism and pointing out its shortcomings \cite{FeynmanHibbs1968}. To answer the questions we should accept the logic of QM in Feynman's formulation for explaining behavior of classical objects; we should revise our views on reality and causality. Following Heisenberg, Fock \cite{Fock1957}, Bohm \cite{Bohm1980}, and Popper \cite{Popper1992}, we should go back to Aristotle's idea about existence as development from possibility into reality and recognize classical determinism as the limiting case of probabilistic (not statistical) causality.

\noindent 

\section{Conclusion}

\noindent 

I assume that the method of path integral, created by R. Feynman for QM, is able to justify and explain the physical sense of some forms of PLA. For this, it is enough to replace the classical representation of objects' motion along a single and unique trajectory by simultaneous motions along an infinite set of possible trajectories or world lines. These motions are described by Feynman's integral over all trajectories.

PLA of classical mechanics can be derived as an approximation from QM laws for scales much larger than Planck's. At the same time, PLA of classical mechanics is an approximation of general relativity for low speeds and weak gravity. In addition, I assume that equations of general relativity could be considered as approximations of the laws of QM when the intricate multidimensional space-time is collapsing into smooth\, 4-dimensional space-time. The axioms of classical and relativistic mechanics can be considered as necessary consequences of QM. As a result, the equations of the main fields of physics could be represented as special cases of the equations of QM. \\

I wish to thank prof. I. Dmitriev for his comments on drafts.  My thanks are due to prof. A. Lipkin and prof. A. Lukyanenko for many useful critical comments and suggestions.

\noindent 
{
\renewcommand{\baselinestretch}{1.0}{\smallskip}
\selectfont
\footnotesize

\bibliographystyle{unsrt}
\bibliography{myreferences}
\smallskip
}

\end{document}